\documentclass[aps,superscriptaddress,prl,showpacs]{revtex4}
\usepackage{epsfig,graphics,amssymb,amsmath,subeqnarray}

\def\dV{\Delta V}

\begin{document}

\title{No many-scallop theorem: Collective locomotion of reciprocal swimmers}
\author{Eric Lauga
\footnote{Email: \texttt{elauga@ucsd.edu}}}
\affiliation{
Department of Mechanical and Aerospace Engineering, University of California San Diego,\\   
9500 Gilman Drive, La Jolla CA 92093-0411, USA.
}
\author{Denis Bartolo
\footnote{Email: \texttt{denis.bartolo@espci.fr}}}
\affiliation{Laboratoire Hydrodynamique et M\'ecanique Physique  PMMH, CNRS UMR 7636, Universit\'es Paris 6 \& Paris 7,  
ESPCI, 10 rue Vauquelin, 75231 Paris Cedex 5, France.
}

\begin{abstract}
To achieve propulsion at low Reynolds number, a swimmer must deform in a way that is not invariant under time-reversal symmetry; this result is known as the scallop theorem. We show here that there is no many-scallop theorem. We demonstrate that two active particles undergoing reciprocal deformations can swim collectively; moreover,  polar particles also experience effective long-range interactions. These results are derived for a minimal dimers model,  and generalized to more complex geometries on the basis of symmetry and scaling arguments. We explain how such cooperative locomotion can be realized experimentally by shaking a collection of soft particles with a homogeneous external field.


\end{abstract}

\date\today
\maketitle

\begin{figure}[t!]
\centering
\includegraphics[width=0.6\columnwidth]{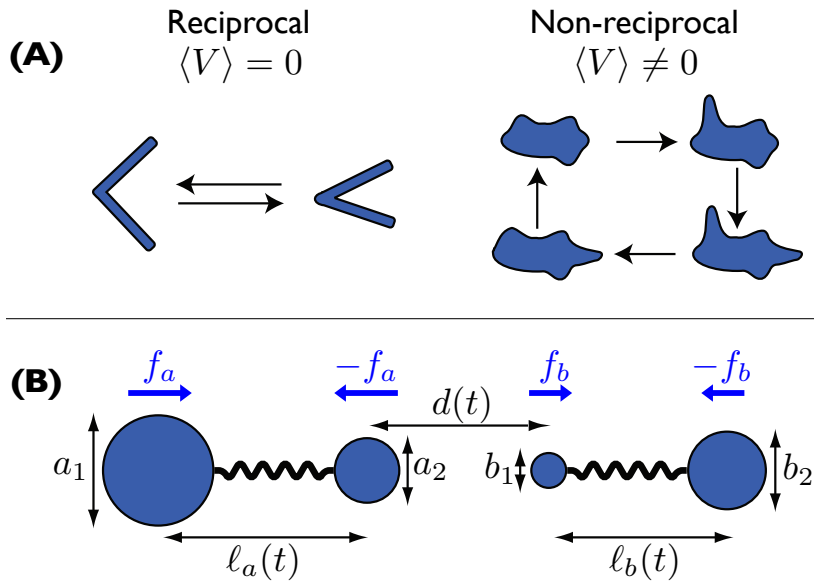}
\caption{(A) A body deforming its shape in a reciprocal fashion, such as a scallop, cannot move on average at low Reynolds numbers  (left, $\langle V \rangle=0$), whereas non-reciprocal deformation leads to net propulsion (right, $\langle V \rangle\neq0$) ; (B) Two force-free dimers  interacting hydrodynamically. The dimers are composed of two solid spheres, of radii $\{a_1,\,a_2\}$ and $\{b_1,\, b_2\}$, have lengths $\ell_a(t)$ and $\ell_b(t)$, and are separated by the distance $d(t)$; the forces on each sphere in a dimer are equal and opposite.}
\label{fig1}
\end{figure}

Microorganisms rely on their ability to swim in order to achieve a variety of biological tasks, including sensing, targeting or feeding~\cite{braybook}. Well-studied examples include bacteria, spermatozoa and ciliated protozoa~\cite{braybook,lighthill76,brennen77}. Given the recent advances in the construction of complex colloidal assembly~\cite{manoharan03, bibette03} and in the coupling of biological machines to artificial microstructures~\cite{atpasescience}, man-made functional micro-swimmers are expected to catch up with real microorganisms \cite{dreyfus05}. 
However, a fundamental challenge in designing artificial micro-swimmers lies in the constraints of the so-called scallop theorem~\cite{purcell77,wilczek87}. Since at small scales, or at low Reynolds number, the Stokes flow equations  are linear and time reversible, swimming can only be achieved by a sequence of shape deformations invariant under time-reversal, or reciprocal; the prototypical reciprocal movement  is that of a scallop which opens and closes its shell (Fig.~\ref{fig1}A, left).  From a design perspective,  the scallop theorem  implies that, in order to move on average, a low-Reynolds number swimmer must have  at least two internal degrees of freedom undergoing a non-reciprocal cycle (Fig.~\ref{fig1}A, right).

In this paper, we show that although an isolated reciprocal active particle cannot move at low Reynolds number, two or more are able to swim collectively; in other words, there is no  many-scallop theorem. We demonstrate that  active particles modulate the relative distance by exploiting hydrodynamic coupling, thereby inducing on each other velocity fields which do not average out to zero and allowing for collective motion to occur. We also show that a set of non-identical reciprocal active particles experience long-range effective interactions which can either be attractive or repulsive depending on the particles geometry. These results are demonstrated  rigorously for a pair of force-free reciprocal dimers, and extended to more general geometries on the basis of symmetry principles and scaling arguments. Experimentally, we show  that  simple elastic particles shaken by a homogeneous oscillating external field can be exploited to obtain collective locomotion.


We consider a collection of prototypical active particles,  force-free dimers,  for which we neglect the flow disturbance created by the links joining them (see Fig.~\ref{fig1}B). 
The i$^{{\rm th}}$ dimer  is composed of two spheres, $i_1$ and $i_2$, separated by the time-varying distance $\ell_i(t)=x_{i_2}-x_{i_1}$, where $x_\alpha$ denotes the position of  sphere $\alpha=i_1,i_2$  along the $x$ axis.
The linearity of Stokes equation implies that the motion of each sphere in a dimer is linearly related to the forces, $f_\beta$,  acting on the spheres from all dimers~\cite{happel}
\begin{equation}
\dot  x_\alpha=\sum_\beta H_{\alpha\beta}f_\beta,
\label{oseen}
\end{equation}
where the $\{H_{\alpha\beta}\}$ are the hydrodynamic mobilities of the spheres ($\beta = j_1,j_2$). We assume that each dimer is force-free,  $f_{i_1}+f_{i_2}=0$, and define $f_i\equiv f_{i_1}=-f_{i_2}$ (Fig.~\ref{fig1}B). The center of the i$^{{\rm th}}$ dimer is denoted $x_i\equiv(x_{i_1}+x_{i_2})/2$. 

\paragraph{Single dimer.} 
We first consider the case of an isolated active dimer, labeled $a$. In that case, we have
$\dot  x_a=H (\ell_a) f_a$ and the dimer elongation satisfies
$\dot  \ell_a =\mu_a(\ell_a)f_a$. Hence, we have $\dot  x_a = H (\ell_a) \dot \ell_a/\mu_a(\ell_a)$, which is an exact derivative and therefore averages to zero over time, $\langle \dot  x_a \rangle=0$, for any periodic sequence of forces or deformations imposed to the dimer. This is the scallop theorem.

\paragraph{Two dimers.} 
We now consider the case of two dimers ($a$ and $b$). 
We demonstrate that, although the dimers are not able to swim when alone, hydrodynamic interactions through the viscous fluid enable collective motion. We restrain our analysis to widely separated dimers in an unbounded fluid; if $d$ is their relative distance, we consider the limit $\ell_i\ll d$ in which case the mobilities in Eq.~\eqref{oseen} are given by the Oseen tensor~\cite{happel}: $H_{\alpha\beta}=1/4\eta \pi d \equiv H(d)$, if $\alpha\neq \beta$, and $H_{\alpha\alpha}=1/6\eta\pi a_\alpha \equiv H(\alpha)$ otherwise ($\eta$ is the fluid viscosity, and $a_\alpha$ the  radius of the  sphere). Performing  a Taylor expansion for the dynamics of the four spheres, Eq.~\ref{oseen}, and keeping only terms of order $1/d^3$ lead to a system of four nonlinear differential equations, two for the positional degrees of freedom, $x_i$, and two for the internal degrees of freedom, $\ell_i$, as given by
\begin{subeqnarray}
\dot x_a&=&M_a f_a- [\partial_x H(x_b-x_a)] \ell_b f_b ,\slabel{eqxa}\\
\dot  x_b&=&M_b f_b+ [\partial_x H(x_b-x_a)] \ell_a f_a \slabel{eqxb},\\
\dot  \ell_a&=&\mu_a(\ell_a) f_a+  \mu_{ab}\ell_a\ell_b f_b\slabel{eqla},\\
\dot  \ell_b&=&\mu_b(\ell_b) f_b+  \mu_{ab} \ell_a\ell_b f_a\slabel{eqlb},
\end{subeqnarray}
where we have defined $M_i = \left[H(i_1)-H(i_2)\right]/2$, $\mu_{ab}=\partial_{xx} H( d) $ and
$\mu_i(\ell_i)=2H(\ell_i)-H(i_1)-H(i_2)$, mobility coefficients associated with the position and the elongation of each dimer respectively. We further restrain our analysis to small amplitude reciprocal motion of each dimer around the time average length, $\bar{\ell_i}\equiv\langle\ell_i\rangle$ (in all that follows brackets stand for time average). More precisely, if we write $\ell_i = \bar \ell_i + O(\epsilon,\epsilon^2)$ and  $f_i =  O(\epsilon,\epsilon^2)$, we keep terms up to $O(\epsilon^2)$ in Eqs.~\eqref{eqxa}-\eqref{eqlb}.
We can then compute the average collective and relative swimming speeds of the dimers, $\langle V\rangle\equiv\langle \dot  x_b + \dot  x_a \rangle /2 $, and $\dV\equiv\langle \dot  x_b - \dot  x_a \rangle $ respectively. To do so, we distinguish the two ways in which the dimers can be physically actuated.

\paragraph{Force-driven motion.}
We first consider the case where the internal forces, $f_i$, are specified. This is analogous to biological swimmers possessing force-generating units (the axoneme for eukaryotic cells~\cite{axoneme}, the rotary motor for bacteria such as {\it E. coli}~\cite{bergmotor}). In that case, we assume the internal force to be known, $O(\epsilon)$, and time-periodic. The force-displacement relation,  Eq.~\eqref{eqla}-\eqref{eqlb}, can then be linearized, and  after some straightforward  algebra, we obtain for a pair of dimers in an unbounded fluid
\begin{subeqnarray}\label{firstresult}
\langle V \rangle & = & \slabel{meanV1}
\left[
\frac{\bar\ell_a (b_2-b_1)}{b_1b_2}
-
\frac{\bar\ell_b  (a_2-a_1)}{a_1a_2}
\right]\frac{\left\langle f_a \int f_b \right\rangle}{48\pi^2\eta^2 d^3}, \\
\dV & = & \left[
\frac{\bar\ell_a (b_2-b_1)}{b_1b_2}
+
\frac{\bar\ell_b (a_2-a_1)}{a_1a_2}
\right]
\frac{\left\langle f_a \int f_b \right\rangle}{24\pi^2\eta^2 d^3}
\cdot\slabel{dV1}
\end{subeqnarray}
Generalization  for an arbitrary response function $H$, including for example boundary effects,  is given in
\footnote{$\{\langle V\rangle,\Delta V/2\}=\mu_{ab}/2\left[\bar \ell_a M_b\{-,+\}\bar \ell_b M_a\right]\left\langle f_a \int f_b \right\rangle$.}.
The results of Eq.~\eqref{firstresult} show that, generically, the scallop theorem breaks down for two active particles interacting hydrodynamically: taken individually, these  particles cannot move, but when interacting through the fluid, they  display collective motion ($\langle V \rangle \neq 0$), and   experience an effective long-range interaction ($\Delta V  \neq 0$), both of which decays in space as $1/d^3$. The direction and sign of the collective and relative speeds depend on the dimers' geometry and actuation; as the simplest example, if we consider sinusoidal forcing of the form $f_i(t)=\bar f_i \cos(\omega t + \phi_i)$, we have $\langle f_a \int f_b \rangle =\bar f_a \bar f_b\sin(\phi_b-\phi_a) / 2\omega$, and locomotion occurs if the two particles are actuated with phase differences,  a direct consequence of the time reversal symmetry of the Stokes equation.

Physically, locomotion of the pair of dimers occurs because their relative distance is oscillating in time, and therefore the flow fields seen by each dimer does not average to zero. In fact, if the dimers are rigidly connected ({\it i.e.} if d(t) is kept constant), we would obtain $\langle V\rangle=0$ at order $1/d^3$, and therefore fluid-mediated forces are the crucial ingredient leading to collective locomotion.

\paragraph{Displacement-driven motion.} 

We now assume the sequence of deformation of each dimer, $\ell_i(t)$, to be specified and time periodic, $\ell_i = \bar \ell_i+ \delta \ell_i(t)$, with $\delta \ell_i(t)=O(\epsilon)$. This is  the relevant limit for (robotic) man-made micro-swimmers. 
We now have to invert the  force-displacement relationship, Eq.~\eqref{eqla}-\eqref{eqlb}, and after some tedious but straightforward algebra we obtain locomotion at speed 
\begin{subeqnarray}\label{secondresult}
\langle V \rangle &=& \left[
\frac{
{\bar\ell_a}(b_2-b_1)
}
{\bar\ell_b(b_1+b_2)}
-
\frac{
{\bar\ell_b}(a_2-a_1)
}
{\bar\ell_a(a_1+a_2)}
\right]
\frac{9\tilde{a}\tilde{b}\langle \delta \ell_b \dot {\delta \ell_a}\rangle}{4d^3},
\quad\quad \slabel{meanV}\\
\dV &=& \left[
\frac{{\bar\ell_a}(b_2-b_1)}
{\bar\ell_b(b_1+b_2)}
+
\frac{{\bar\ell_b}(a_2-a_1)}
{\bar\ell_a(a_1+a_2)}
\right]
\frac{9\tilde{a}\tilde{b} \langle \delta \ell_b \dot {\delta \ell_a}\rangle}{2d^3}
\slabel{dV},
\end{subeqnarray}
where we have defined $\tilde{i}=i_1i_2/(i_1+i_2)$ ($i=a,b$).
The general expressions for an arbitrary kernel $H$ are given in 
\footnote{
$\left\{\langle V\rangle,\frac{\Delta V}{2}\right\}= \frac{\bar \ell_a \bar \ell_b\mu_{ab}}{2\mu_a\mu_b} 
\left[\frac{ M_b}{\mu_b} \frac{\partial \mu_b}{ \partial\ell_b} \{-,+\}\frac{ M_a}{\mu_a}\frac{\partial \mu_a}{ \partial\ell_a} \right]\langle \delta \ell_b \dot {\delta \ell_a}\rangle$.
}. 
Similarly to  force-driven motion, the scallop theorem does not hold for a pair of reciprocal bodies, and locomotion  arises if there is a non-zero phase difference between the deformation of each particle. If  the dimers were rigidly connected, we would  obtain in this case a $O(1/d^3)$ swimming velocity, but with a  sign opposite to  that  arising from hydrodynamic interactions, Eq.~\eqref{meanV}~\footnote{$\langle V \rangle =
\left[\bar\ell_a\left(\frac{b_1-b_2}{b_1b_2}\right)-{\bar\ell_b\left(\frac{a_1-a_2}{a_1a_2}\right)}\right]
{3 \widetilde{ab}\langle \delta \ell_b \dot {\delta \ell_a}\rangle}/{4d^3}$.}.

\begin{figure}[t!]
\centering
\includegraphics[width=0.7\columnwidth]{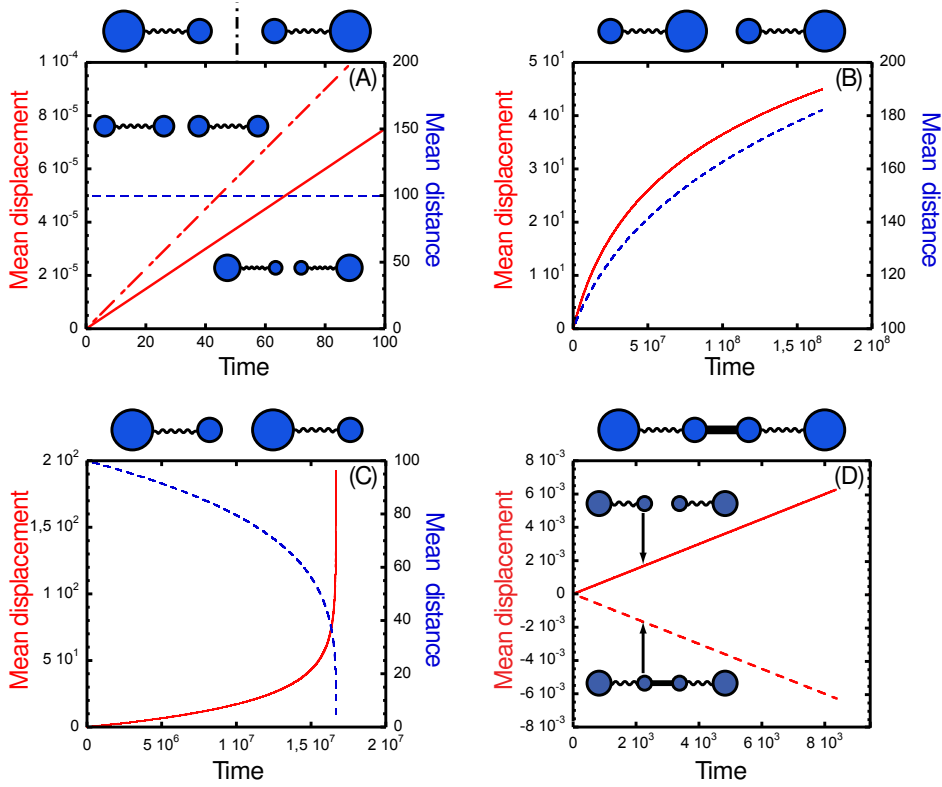}
\caption{Dynamics of identical dimers interacting hydrodynamically. 
Units are chosen so that $\delta \ell_a(t)=\cos(t)$, $\delta \ell_b(t)=\sin(t)$.
A: mean displacement, $\langle \int  V\,dt \rangle $, for two mirror-images polar dimers ($a_1\neq a_2$, solid line); mean displacement for two mirror-images apolar  dimers ($a_1=a_2$, dash-dotted line); mean distance, $\int\Delta V\,dt$, between the two dimers (polar and apolar cases,  dashed line). 
B: mean displacement (solid line) and mean separation distance (dashed line) for two identical polar dimers pointing in the same direction. C: same as \ref{fig2}B but for dimers pointing in the opposite direction; note the change in the sign of the dimer-dimer effective interaction. 
D: comparison between the swimming speed of two active dimers and a single swimmer made of two rigidly connected dimers (constant $d$). 
In all figures, the lengths $\ell_a$ and $\ell_b$ oscillate with the same amplitude and frequencies, and  a relative phase of $\pi/2$.  In the chosen units, $\bar{\ell}_a=\bar{\ell}_b=10$, $d|_{t=0}=100$, $a_1=3$, $a_2=3/2$ for the polar dimers and $a_1=a_2=2$ in the apolar case.}   
\label{fig2}
\end{figure}

\paragraph{Identical swimmers.} 
In the particular case where the two dimers are identical ($a_i=b_i$, $i=1,2$, and $\bar \ell_a=\bar \ell_b\equiv \ell$), the result of Eqs.~\eqref{meanV1}-\eqref{meanV}  cancel out. 
One needs to go to higher order in the asymptotic expansions to obtain the swimming kinematics, which  is readily done, and locomotion is obtained at order $1/d^4$, with velocities
\begin{equation}
\langle V\rangle 
= \frac{\ell^2 \left\langle f_a\int f_b \right\rangle}{16\pi^2\eta^2 \tilde{a}d^4},\,\, \text{and}\,\,\,
\langle V\rangle 
= \frac{9 \tilde{a} \ell^2 \langle\delta \ell_a\delta \dot{\ell}_{b}\rangle}{4d^4},
\label{d4}
\end{equation}
for force- and displacement-driven motion  respectively.

In Fig.~\ref{fig2}A-C we highlight  the main features of the collective dynamics of a pair of identical dimers in the case of displacement-driven motion, by integrating numerically Eqs.~\eqref{dV} and \eqref{d4}; results for force-driven motion are similar. We first show in  Fig.~\ref{fig2}A  that, on average, mirror-symmetric swimmers remain at the same relative distance and swim collectively with a constant velocity.  
When the swimmers do not display mirror-image symmetry, we show in Fig.~\ref{fig2}B-C that they undergo a repulsive or an attractive effective interaction depending on their relative orientation. 
The  dimers are separating/approaching at rate $\dot d \sim \pm  1/d^3$, and therefore $d\sim (t_0 \pm t)^{1/4}$; accordingly, the mean speed decreases as $\langle V \rangle \sim  (t_0 \pm t)^{-1}$, and the distance traveled by the swimmers can be arbitrarily (logarithmically) large, $\int \langle V \rangle  \,dt \sim \log (t_0 \pm t) $ \footnote{If the dimers are not identical, we have the scaling $\langle V \rangle \sim 1/d^3$, resulting in $\int \langle V \rangle  \,dt \sim t^{1/4}$ in the case of separating swimmers and  $\int \langle V \rangle  \,dt \sim O(1)$ otherwise.}.
Note that in the case of relative attraction, a proper description of the near-contact hydrodynamics would regularize the finite-time  singularity displayed in Fig.~\ref{fig2}B; in an experiment,  short-range surface forces (such as van der Waals) would lead to self-assembly of the two dimers into a four-spheres non-reciprocal swimmer. 

\paragraph{Generalization.}
The results above demonstrate the emergence of collective motion and of long-range interactions between two dimers embedded in a viscous fluid, and can be generalized in a number of ways.

First, our calculations  were made under the assumption of an infinite fluid, but the results remain valid in other geometrical settings, such as a Hele-shaw cell, or in confined geometries  typical of microfluidic applications, by changing appropriately the Oseen-like kernel $H(d)$. In addition, our dimers are made of two connected spheres, but other solid bodies with $l/a\ll 1$  can be considered by simply modifying the values of $H(\alpha)$ [21,22].

Furthermore,  Eqs.~\eqref{firstresult} and~\eqref{secondresult} also apply for stochastic fluctuations of the length of each dimer, and any correlation between the noisy shaking sources acting on each dimer is seen to  lead to non-zero swimming velocities and effective interactions between the two active objects. 

Theoretically,  the $1/d^3$ spatial decay of the velocities with the inter-dimer distance in Eqs.~\eqref{firstresult} and~\eqref{secondresult} could have been anticipated since the locomotion arises from the rectification of interacting force dipoles decaying as $1/x^2$ at long distance~\cite{pooley07}; this scaling argument also hold for any pair of active particles having a single deformation degree of freedom aligned with the $x$-axis. In the small deformations limit, their collective and relative swimming velocities is expected to be of the generic form:
\begin{eqnarray}
\langle V\rangle&\sim&{\textstyle \langle f_a\int f_b\rangle}\sum_{n\geq3}\frac{\alpha_n}{d^n}, \,\,\,\,
\Delta V \sim{\textstyle \langle f_a\int f_b\rangle}\sum_{n\geq3}\frac{\gamma_n}{d^n},\,\,\,\\
\langle V\rangle & \sim &  \langle\delta \ell_b\delta \dot{\ell_a}\rangle\sum_{n\geq3}\frac{\beta_n}{d^n},\,\,\,\,
\Delta V \sim \langle\delta \ell_b\delta \dot{\ell_a}\rangle\sum_{n\geq3}\frac{\delta_n}{d^n},\,\,\,
\end{eqnarray}
for force- and displacement-driven motion respectively, and 
where $\{\alpha_n,\beta_n,\gamma_n,\delta_n\}$  depend solely on the shape of the  particles \footnote{{Linear and quadratic terms proportional to $\delta \dot{\ell_i}$ and $\delta \ell_i\delta\dot{\ell_i}$ have a zero mean. Thus, at lowest order, the only non-zero term has the antisymmetric form $\langle \delta \ell_b\delta \dot{\ell_a}-\delta \ell_a\delta \dot{\ell_b}\rangle =2\langle \delta \ell_b\delta \dot{\ell_a}\rangle $ for reciprocal deformations sequences~\cite{raminarmand}.}}. 
Note from Eqs.~\eqref{dV1}-\eqref{dV}   that for two mirror-image particles, $\gamma_3=\delta_3=0$;  this result is actually true at all orders, and two swimmer with mirror-image symmetry verify $\Delta V=0$. 
Indeed the flow and pressure fields induced by the beating of two mirror-images dimer are invariant under the combination of time-reversal and mirror symmetries. Consequently,  $\Delta V$, which has the symmetry of a velocity gradient, transforms into $-\Delta V$, and therefore $\Delta V=0$. Similarly, we have $\alpha_3=\beta_3=0$ for identical dimers; this is a consequence of the two (rectified) dipoles having opposite contributions on each dimer, and is generally true for a pair of identical particles.

Beyond the simple one-dimensional models considered here,  all types of reciprocal motion (including those with three-dimensional shape deformation) are also expected to display collective motion induced by hydrodynamic interactions. And although we have emphasized  locomotion in this paper,  our results can be extended to the dual problem of pumping fluid by anchored bodies.

\paragraph{Soft swimmers.} We now turn to the experimental realization of these ideas.  Actuating a collection of  active particles with out-of-phase conformational changes is difficult, and it would be preferable to devise a framework where an homogeneous external field could produce directed locomotion.
We show below that this is possible if the particle are soft: indeed, since soft particles in a viscous fluid posses relaxation times, different particles will naturally react to an external shaking with phase differences, and locomotion will ensue. 
Practical examples of soft-particle actuation include  bending of magnetic filaments by magnetic fields \cite{dreyfus05},  temperature or light actuation of  liquid-crystal elastomers \cite{elastomer,elastomer2}, or self-sustained chemical reactions for swelling of gels \cite{gels}. 

We thus consider a pair of force-free  apolar and identical  dimers (radius $a$, average length $\ell$) subject to a homogeneous external shaking. We write $f_i=f^{\rm shake}+f_i^{\rm relax}$, for $i=a,b$, where the force $f^{\rm shake}$ is externally produced and the same for each dimer (homogeneous forcing). 
The force $f_i^{\rm relax}$ is the internal (elastic) response of the dimer, and write  $f_i^{\rm relax}=k_i \delta \ell_ i $ where $k_i$ is the dimer stiffness; in that case, the intrinsic dynamics of each active particle is characterized by a relaxation time scale $\tau_i = {3\pi\eta a}/{k_i}$.
Assuming a monochromatic shaking $f^{\rm shake}=f_0\cos(\omega t)$ for simplicity,  integration of Eqs.~\eqref{eqla}-\eqref{eqlb} leads to collective motion  with speed
\begin{equation}\label{visco}
\langle V \rangle = 
\left[\frac{(\tau_a-\tau_b)(\omega^2\tau_a\tau_b)}{(1+\omega^2 \tau_a^2)(1+\omega^2 \tau_b^2)}\right]
\frac{\ell^2 f_0^2}{16\pi^2\eta^2 d^4 a}  ,
\end{equation}
and $\Delta V=0$ by symmetry. We see that,  under homogeneous forcing, the only condition necessary to obtain locomotion is  $\tau_a \neq \tau_b$.
Locomotion always occurs in the direction of decreasing relaxation time, {\it i.e.} the stiff dimer is pulling the soft one; optimal locomotion occurs when the system is actuated with frequency $\omega \sim (\tau_a\tau_b)^{-1/2}$ and when the ratio of relaxation times is large.

A final relevant  example  is the case where one dimer is purely passive, {\it i.e.} 
its elongation is not coupled  to the external field.
Specifically, we assume dimer  $a$ to be actively shaken, with $f_a^{\rm shake}=f_0\cos(\omega t)$, while  dimer $b$ is passive, and $f_b^{\rm shake}=0$. In that case, the beating of the passive dimer arises from the hydrodynamic interactions with the active dimer, and its amplitude is a function of the distance $d$. Collective motion arises with speed
\begin{equation}\label{passive}
\langle V \rangle =-\left[\frac{ \omega^2 \tau_a\tau_b}{(1+\omega^2\tau_a^2)(1+\omega^2\tau_b^2)}\right]\frac{3\tau_a\ell^4f_0^2}{32\pi^2 \eta^2 d^7},
\end{equation}
and the active particle is seen to ``pull'' the passive one.

\paragraph{Perspective.} 
We have shown in this paper that there is no many-scallop theorem: two bodies with reciprocal deformation can exploit hydrodynamic interactions to obtain collective locomotion and effective long-range interactions. 
On the basis of our two-body investigation, the  generalization to a large number of 
reciprocal soft active particles interacting through a fluid is  expected to display both spatial 
and spatio-temporal  spontaneous symmetry-breaking (coarsening, ordering and directed large-scale motions). Similar cooperative effects for coupled active (but non-self propelled) particles have already been demonstrated in the very different context of molecular motors~\cite{molecular_motors}, active membranes dynamics~\cite{active_membrane} and vibrated nematic grains~\cite{vibrated_nematic}.

This work was supported in part by the NSF (grant CTS-0624830 to EL), by ESPCI and by Denis Diderot University (DB).

\bibliography{many_scallop}
\end{document}